\documentclass[10pt,journal,twocolumn,twoside]{IEEEtran} % formal style
\usepackage{xpatch}
\usepackage{graphicx}
\usepackage{epstopdf}
\usepackage{float}
\usepackage{algorithmic}
\usepackage{array}
\usepackage{amsmath}
\usepackage{amssymb}
\usepackage{mdwmath}
\usepackage{mdwtab}
\usepackage{eqparbox}
\usepackage{stfloats}
\usepackage{fixltx2e}
\usepackage{tabularx}
\usepackage{cases} 
\usepackage{booktabs}
\usepackage{flushend}
\usepackage{threeparttable}
\usepackage{xcolor}
\usepackage{makecell}

\usepackage[boxed,ruled,commentsnumbered]{algorithm2e}
\usepackage{url}
\usepackage{cite}
\ifCLASSOPTIONcompsoc
\usepackage[caption=false,font=normalsize,labelfont=sf,textfont=sf]{subfig}
\else
\usepackage[caption=false,font=footnotesize]{subfig}
\fi
\allowdisplaybreaks[4]

\makeatletter

\newcommand{\Rmnum}[1]{\expandafter\@slowromancap\romannumeral #1@}
\makeatother

\makeatletter
\renewcommand*{\@opargbegintheorem}[3]{\trivlist
      \item[\hskip \labelsep{\bfseries #1\ #2}] \textbf{(#3):}\ }
\makeatother

\begin{document}

\makeatletter
\def\changeBibColor#1{%
  \in@{#1}{}%  list of colored bib items
  \ifin@\color{red}\else\normalcolor\fi
}
 
\xpatchcmd\@bibitem
  {\item}
  {\changeBibColor{#1}\item}
  {}{\fail}
 
\xpatchcmd\@lbibitem
  {\item}
  {\changeBibColor{#2}\item}
  {}{\fail}
\makeatother

\title
{Emergent Semantic Communications for Mobile Augmented Reality: Basic Ideas and Opportunities}
\author{Ruxiao Chen, and Shuaishuai Guo,~\IEEEmembership{Senior Member, IEEE} 

\thanks{R. Chen and S. Guo are all with the School of Control Science and Engineering, Shandong University, China. S. Guo is also with Shandong Key Laboratory of Wireless Communication Technologies, Shandong University, China (e-mail: ruxiaochen333@gmail.com; shuaishuai$\_$guo@sdu.edu.cn). }
   }
\maketitle

%\newpage

\begin{abstract} 
Mobile augmented reality (MAR) is widely acknowledged as one of the ubiquitous interfaces to the digital twin and Metaverse, demanding unparalleled levels of latency, computational power, and energy efficiency. The existing solutions for realizing MAR combine multiple technologies like edge, cloud computing, and fifth-generation (5G) networks. However, the inherent communication latency of visual data imposes apparent limitations on the quality of experience (QoE). To address the challenge, we propose an emergent semantic communication framework to learn the communication protocols in MAR. Specifically, we train two agents through a modified Lewis signaling game to emerge a discrete communication protocol spontaneously. Based on this protocol, two agents can communicate about the abstract idea of visual data through messages with extremely small data sizes in a noisy channel, which leads to message errors. To better simulate real-world scenarios, we incorporate channel uncertainty into our training process. Experiments have shown that the proposed scheme has better generalization on unseen objects than traditional object recognition used in MAR and can effectively enhance communication efficiency through the utilization of small-size messages.

\end{abstract}

% \begin{IEEEkeywords}
% Mobile augmented reality(MAR), emergent semantic communications, Lewis signaling game
% \end{IEEEkeywords}

\section{Introduction} 
\IEEEPARstart{A}{s} one of the primary interfaces to the digital twin and Metaverse, mobile augmented reality (MAR) enables users to blend digital content seamlessly with the real world. It allows individuals to view and interact with virtual objects and information overlaid on their physical surroundings through the device's camera and display. The realization of MAR tasks often involves processing computationally intensive visual data. The underlying technologies like the fifth generation (5G) networks and mobile edge computing have accelerated the implementations of MAR. However, MAR's requirements on latency, mobility, and endurance are stringent, and with numerous pictures to process per frame, the conventional data-oriented communication method is overwhelmingly burdensome. Data-oriented communication tends to transmit the symbols representing the original data, which has been well-studied and realized. However, this level of communication is approaching the physical constraints of the underlying infrastructure \cite{Luo2022}. Semantic communication, on the other hand, goes beyond the mere transmission of data by focusing on comprehending the essence and concepts embedded within the information and has been gaining increasing attention \cite{GuoSept.2022,Guo2023}. Due to the above features, semantic communication is promising for MAR tasks that involve handling extensive raw data and require significant computational power.

A large body of artificial intelligence (AI)-based algorithms for processing video streams were proposed for MAR tasks. For example, Ren \emph{et al.} proposed a motion-aware scheduler \cite{Ren2022} to select the keyframe from the video source for offloading, thereby improving the computational efficiency of the edge system. In \cite{Al-Shuwaili2017}, Shuwaili \emph{et al.} leveraged the inherent collaborative nature of MAR that mobile devices connected to the same base station have partly shared inputs and outputs video stream to avoid extra computing, then using a successive convex approximation method (SCA) to solve the non-convex optimization problem. In\cite{Lee2020}, Lee \emph{et al.} proposed a reinforcement learning-based server-client controlling scheme that conducts class-wise characteristic analysis from the experience so that it could control the MAR service quality adaptively.

The aforementioned works primarily focus on the first level of communication, specifically centered on transmitting the original data. To the best of our knowledge, no existing research has explored the utilization of semantic-level communication in MAR. For video processing tasks like object recognition in MAR scenarios, existing methods commonly employ two approaches. The first involves utilizing a convolutional neural network (CNN) to extract feature vectors, which are then transmitted to servers for subsequent computation \cite{Chen2023}. The second approach entails directly transmitting the original video stream to servers for processing. Regardless of the method, the amount of data transferred is enormous. In specific scenarios, it’s apparent that not all video features are required. Instead, only specific segments of the video hold relevance \cite{Seo2021}. For instance, the usage case of MAR we analyzed in this paper focused on bird recognition and displaying related information to the user, extraneous objects such as trees or people need not be considered, and to process all these videos without selection is a huge waste of computation. By employing emergent communication for video transmission, we can effectively encapsulate each video frame's concept and abstract ideas within a concise set of messages \cite{Dessi2021}. These messages facilitate communication efficiency between user devices and servers. Notably, the data size of these messages is orders of magnitude different from the two previously described methods.

To accomplish this objective, our system employs two intelligent agents: the speaker and the listener, which are trained via a modified Lewis signaling game \cite{Lewis1970}. In this game, the speaker is tasked with generating a message based on a concept, utilizing stored pictures of the concept. Subsequently, the listener uses this message to determine which pictures correspond to the given concept. To some extent, this approach leverages the memory of stored concept pictures, trading off communication time and computation, which proves to be useful by experimental evidence when employed in MAR tasks. Besides, the emergent communication protocol can be generalized to concepts unseen in the training process\cite{Mu2021}. For instance, if blue squares and green circles are seen while training, the generated messages are able to describe blue circles to the listener. Traditional object recognition like YOLO\cite{Wang2022} is unable to generalize to this level. Additionally, our training process takes into account channel uncertainty, which is a more complete simulation of the real world \cite{Kucinski2021}. Experimental evidence has demonstrated it improves the robustness and flexibility of the system when encountering different channels.

\begin{figure*}
  \centering
  \includegraphics[width=18cm]{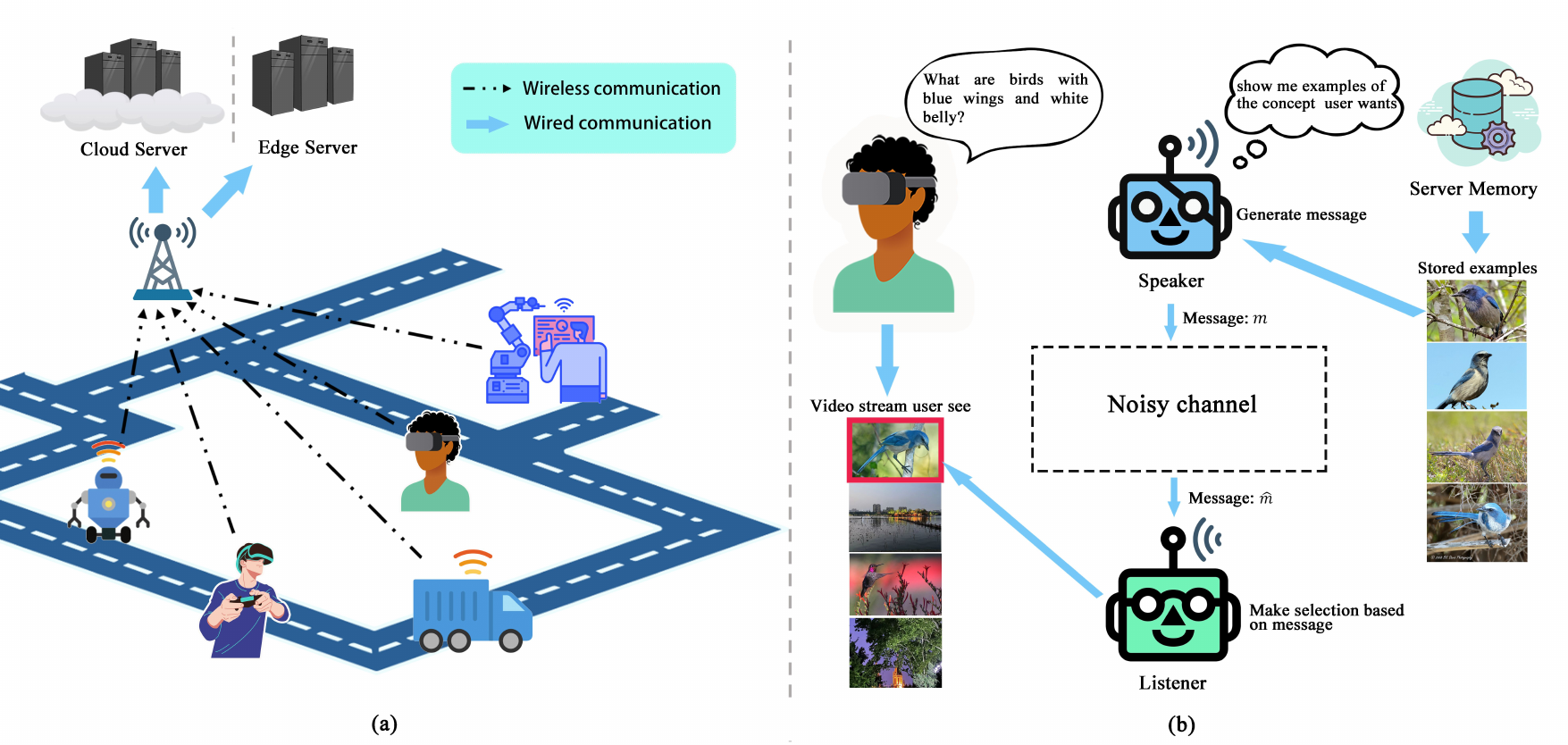}
  \caption{System architecture (a), users located in the same area share the same base station. System application scenario (b), a MAR-aided birds education system for students. } 
  \label{fig1}
\end{figure*}

Drawing all the insights above, this article proposed an emergent semantic communication framework for MAR. The main contribution of this article can be summarized as follow:
\begin{itemize}
	\item Establishing a basic idea and structure for implementing emergent communication in the field of MAR for visual data processing.
        \item Demonstrating the emergent semantic communication's superior generalization ability and significantly smaller communication data sizes, identifying its potential benefits in industrial, societal, and business applications.
	\item Considering the channel uncertainty in real-world scenarios in the training process of emergent communication, we achieve increased robustness when confronted with varying channels.
\end{itemize}

\section{System Architecture}

We consider a typical MAR application of education, as depicted in Fig. \ref{fig1} (a). The system encompasses a substantial user population within a specific area, all of whom are connected to a base station. The base station, in turn, establishes a connection with multiple edge and cloud servers. During operation, the users' devices offload their computational tasks to the base station, continuing their execution on one of the servers. And the results of the computation will then be transmitted back to users’ devices for rendering or other functions. We examine a specific education-oriented scenario illustrated in Fig. \ref{fig1} (b). Mobile devices are employed to capture desired types of birds within the video stream according to users’ requirements. Subsequently, these devices seamlessly overlay pertinent information about the identified birds on the original video. 

When encountering an unfamiliar bird, users can describe it based on features such as the color of its wings, the shape of its bills, etc. These descriptions can be regarded as concepts that can be mathematically expressed and transmitted to speakers located on servers. Examples of all concepts are stored in the server memory in the first place, when a concept is required, the speaker can extract the specific examples of the required concept for learning. After learning, the speaker can summarize the concept into a message, denoted as $m$, which is a discrete sequence of length $L$ with a vocabulary size of $V$. The message $m$ is then transmitted to the listener on the user’s device. It’s worth noting that during transmission, the message will experience interference from channel uncertainty, the training process should take this into consideration. The message after undergoing the channel uncertainty can be represented as $\hat{m}$. Utilizing the message $\hat{m}$, listeners are able to select the relevant video stream given by the AR device that belongs to the concept user required. Since the emergent communication channel is distributed, the channel uncertainty can be realized and represented using error rate $\epsilon$. It refers to the likelihood of the current character in a message undergoing a transformation into a different character.

To further illustrate the problem at hand, we now conduct a detailed modeling analysis for the aforementioned MAR application case. It can be divided into five interconnected components: a video capturer, a feature extractor, a mapper, a tracker, an object recognizer, and a render. The video capturer is a camera capturing the raw video frame, which is then processed by the feature extractor to extract its feature points. The mapper leverages the feature points to construct a digital model of the 3-dimensional (3D) environment, while the object recognizer employs the feature points to identify specific objects. The tracker is responsible for tracking the identified object across subsequent frames. Finally, the render component combines all the positional and image information, enabling the overlaying of virtual content onto the original video, such as the introduction of the birds. 

In this application case of MAR, the main consumption of computation resources and transmission bandwidth is the object recognition component. Thus, this component is where emergent semantic communication can be implemented to improve the quality of service. The prevailing approaches commonly adopt YOLO \cite{Wang2022} to achieve this goal: to use CNN to extract the feature vectors of each video stream for further computation. As illustrated above, these methods typically involve extracting the feature vectors beforehand and transmitting them to the server or directly sending the entire video. However, despite their undeniable accuracy, both approaches suffer from the challenge of substantial data size for transmission and demanding computational requirements.

% \begin{figure}
%   \centering
%   \includegraphics[width=8.5 cm]{fig/AR.jpg}
%   \caption{Main components of MAR education application and their interdependency.} 
%   \label{fig2}
% \end{figure}

\section{Basic Ideas of Emergent Semantic Communications} 
\begin{figure}
  \centering
  \includegraphics[width=9 cm]{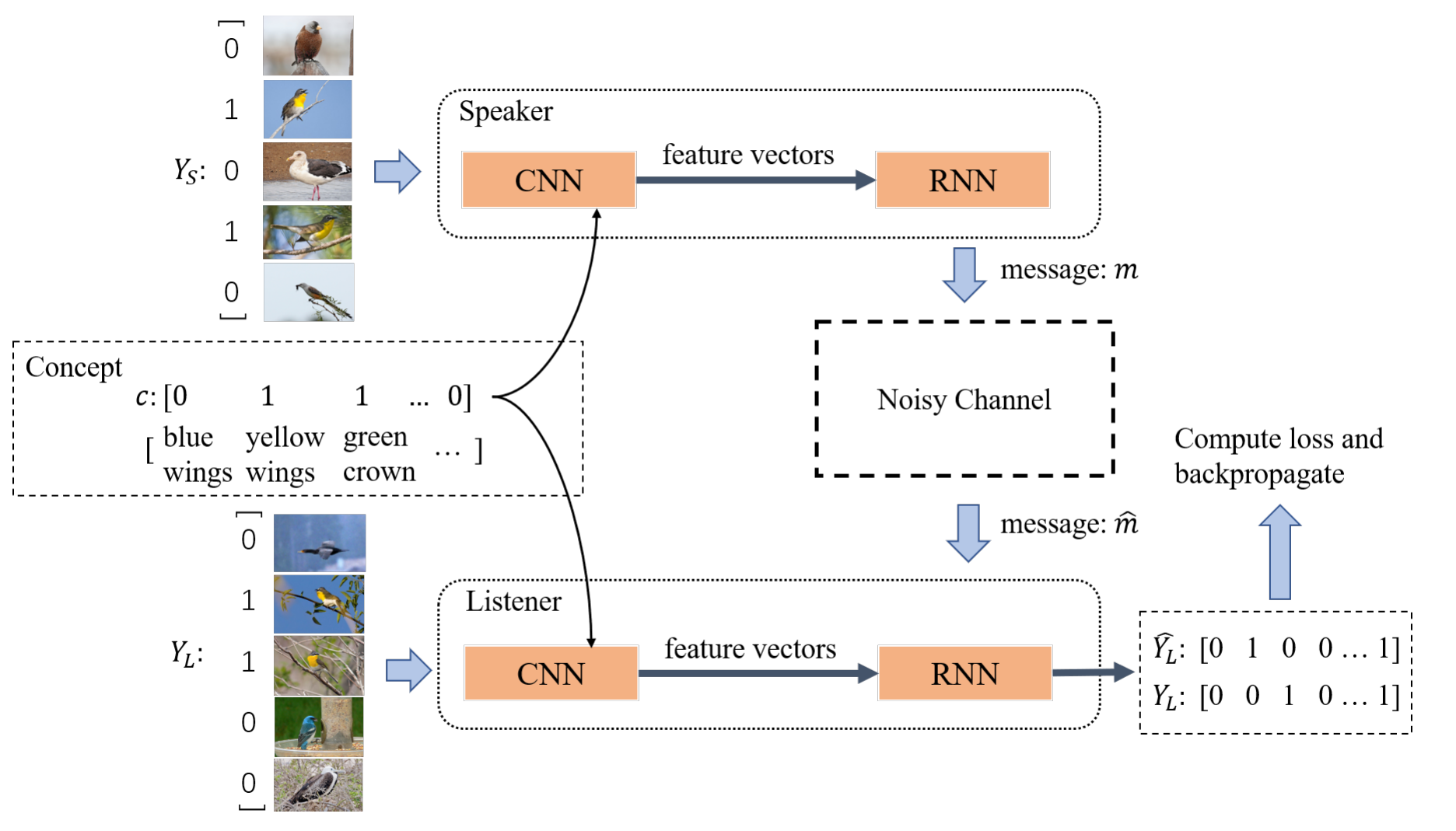}
  \caption{The training process of a modified Lewis signaling game that focuses on concept identification.} 
  \label{fig3}
\end{figure}

To make the communication protocol emerge, the training process of our system implements a modified Lewis signaling game, as depicted in Fig. \ref{fig3}. Lewis signaling game \cite{Lewis1970} is widely used in the training of emergent communication that contains two intelligent agents, the speaker, and the listener. In this game, the speaker is given a target picture to generate a discrete message, and the listener is given a set of pictures including the target picture and several distracting pictures. Based on the given message, the listener’s goal is to successfully identify the target picture from other distracting pictures. 

In our game setting, however, the focus is not on specific objects but rather on the concept of an object. In our definition, a concept consists of several attributes, and each attribute can have different values \cite{Ohmer2022}. For example, a concept $c$ can have two attributes: the color of wings and the shape of bills. The values of the attribute color of wings can be green and blue, and the values of the attribute shape of bills can be dagger and hooked.

Given a concept and a set of objects belonging to that concept, along with other distracting objects, the speaker generates a message that conveys the abstract idea of the concept based on its observation. The listener then uses this message to distinguish the pictures belonging to the concept from the distracting ones. In order to enhance generalization capability, the pictures provided to the listener and speaker are different, despite belonging to the same concept. The speaker and the listener contain a CNN and a recurrent neural network (RNN) respectively. The CNN is responsible for extracting the feature vectors of a given picture and RNN is responsible for encoding or decoding messages. A concept $c$ can be presented as a binary matrix, with 0 meaning the absence of an attribute, and 1 meaning the presence of an attribute. The actual value $Y_L$ and the prediction $\hat{Y_L}$ can be represented as a binary matrix with 0 meaning distracting picture and 1 meaning target picture. To fine-tune their neural network parameters, the two agents continuously adjust them based on the disparity between the prediction and the actual value. 

The message data size is typically small, as it is a discrete sequence of length $L$ with a vocabulary size of $V$. Thus, a single-bit error caused by channel uncertainty might lead to a substantial misinterpretation of the final outcome. To account for the influence of channel uncertainty, we integrate different message error rates into both the training and simulation process. 

\section{Evaluation} 

The experiments were conducted in a simulated MAR system, with video processing tasks generated every time interval. During the training process, we set the message length $L$ to $4$ and the maximum vocabulary size $V$ to $14$. The training utilized the Caltech-UCSD Birds dataset\footnote{http://www.vision.caltech.edu/datasets/cub\_200\_2011/}, and we trained the model for 100 epochs to ensure comprehensive learning.

\begin{figure}
  \centering
  \includegraphics[width=9.5cm]{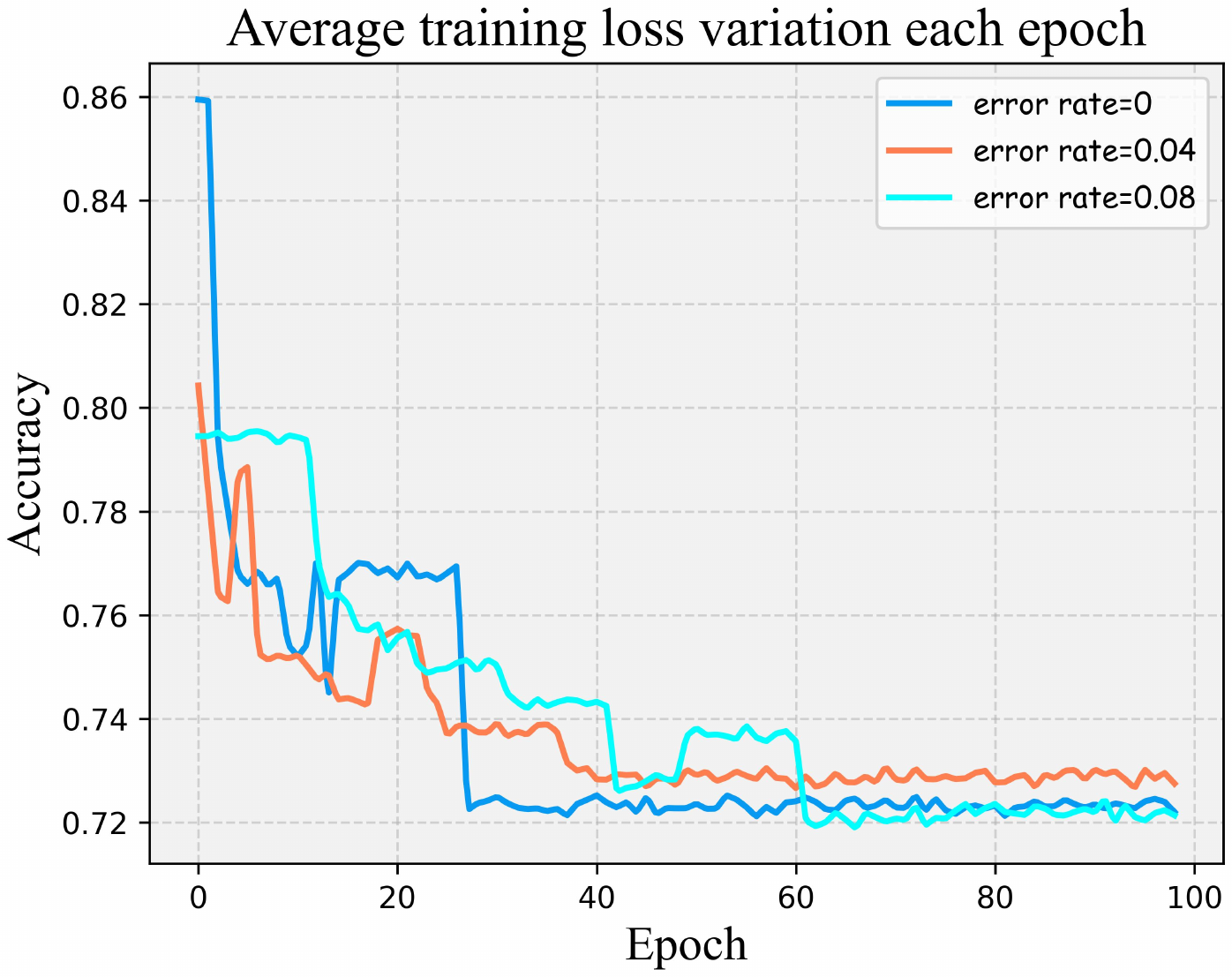}
  \caption{Average training loss for each epoch} 
  \label{fig4}
\end{figure}

We simulate the real channel environment as a discrete message error channel.  We investigate different error rates in the channel from $0$ to $0.10$. Our first finding is a higher error rate leads to slower convergence, as demonstrated in Fig. \ref{fig4}. For example, when the error rate was $0$, the system converged at epoch $29$, while at an error rate of $0.08$, convergence occurred at epoch $62$. This observation suggests that an appropriate error rate enables the neural network to explore a wider range of possibilities, preventing it from becoming trapped in a local optimum solution.

\begin{figure}
  \centering
  \includegraphics[width=9.5cm]{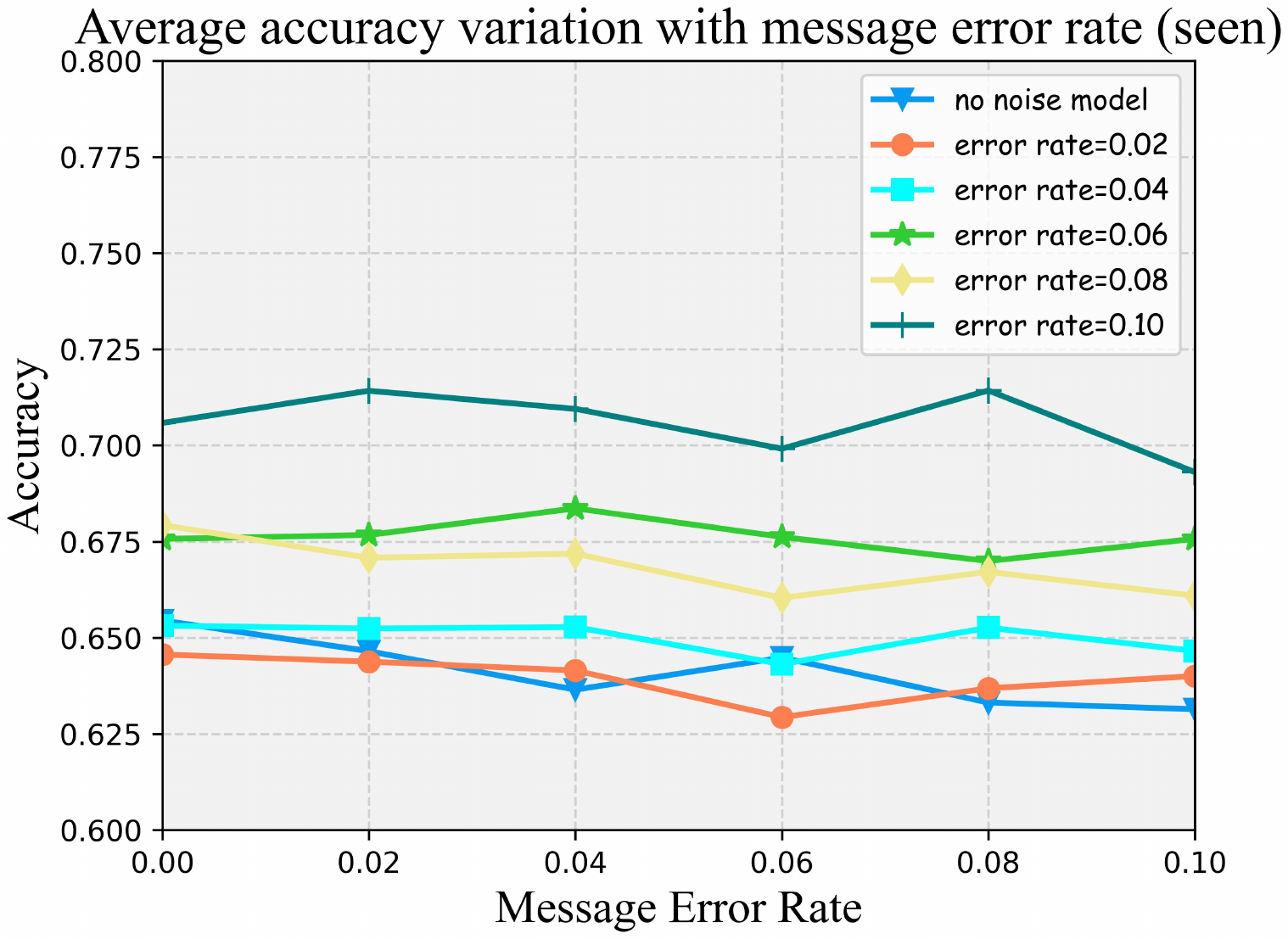}
  \caption{Average accuracy for each model on the seen concept} 
  \label{fig5}
\end{figure}

We first test the accuracy of our trained model on the concept seen during training at different message error rates, as shown in Fig. \ref{fig5}. The identification accuracy of the model that didn't consider message errors while training declined as the message error rate averages rise. The results from the models with error rates set to $0.02$ and $0.04$ suggest that a moderate error rate can enhance the system's robustness in the presence of message errors. The lines of these two models exhibited more stability compared to the model without message errors as the message error rate increased. Interestingly, increasing the message error rate beyond a certain point not only enhances the stability of the system but also improves the accuracy rate. This coincides with previous studies done by Kucinski \emph{et al.}\cite{Kucinski2021}. This outcome can be attributed to the random shuffling of the message, which facilitates the model in exploring a wider range of possibilities.

\begin{figure}
  \centering
  \includegraphics[width=8cm]{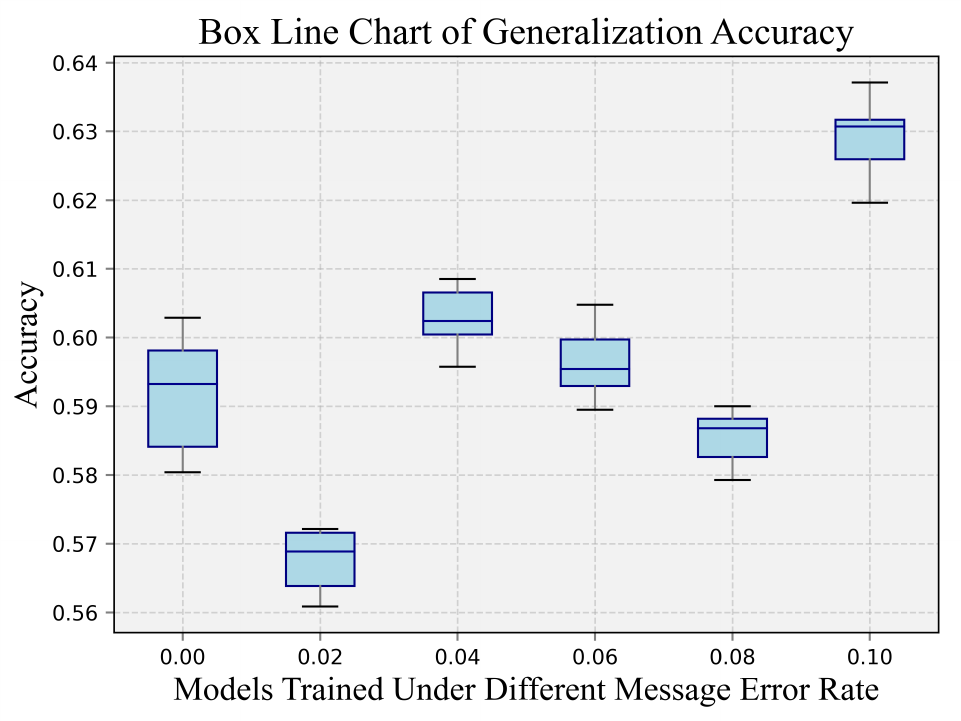}
  \caption{Average accuracy for each model on the unseen concept} 
  \label{fig6}
\end{figure}

Subsequently, we evaluate the generalization capability of the trained model by examining its performance on unseen concepts at different message error rates. For the sake of comprehension and maximum utilization of the message sizes, the ideal model is each character in the message represents an attribute \cite{Ohmer2022}. This feature of communication protocols enables it to arrange and combine different attributes which is why it can generalize to unseen concepts. The obtained results Fig. \ref{fig6} demonstrate a certain degree of generalization to unseen concepts; however, the accuracy achieved is still suboptimal.

To better illustrate the communication efficiency of our model, we conduct a theoretical comparison between our system and the two other approaches mentioned above. The size of the pictures used in the training and evaluation ranged from $70$ kilobytes to $100$ kilobytes. The feature vector extracted by the CNN network of YOLOv5 is about 384 bytes\cite{Wang2022}. For the message length and vocabulary used in the system, as there are $14$ possibilities for each character, each character can be represented by a one-hot vector of length $14$, which is $14$ bits. The data size of messages with four characters is thus $4\times 14=56$ bits. The data size that needs to be transmitted for each video frame is listed in Table 1. 

For mainstream target detection algorithms like YOLO, the accuracy of the latest version YOLOv7-E6 has an average precision of $73.5$\% at a frame rate of $56$ fps \cite{Wang2022}. Although our proposed system can't compete with it in accuracy, communication efficiency and generalization ability to unseen concepts make it a promising solution in the context of AR-oriented tasks. Besides, the structure of our neural network is relatively small compared to that of YOLO. Recent researches indicate that increasing the scale of emergent communication training\cite{Chaabouni2022} or introducing a more rigorous environment may yield additional benefits in generating an ideal communication protocol, and thus are left to our future work.

\begin{table}
\centering
\caption{Data Size Comparison Among Three Different Methods} %
\begin{tabular}{ccc} 
\toprule 
Original video stream & Feature vectors & Emergent messages \\
\midrule 
70 - 100 kilobytes & 384 bytes & 56 bits \\
\bottomrule 
\end{tabular}
% \begin{tablenotes}[flushleft] % 添加注释
% \item[] Tabel 1: This is a note below the table.
% \end{tablenotes}
\end{table}

\section{Opportunity and Challenges}
Implementing emergent semantic communication in the field of MAR is an unexplored area before, and it presents untapped potential when it comes to visual data. For example, vehicles can extract the abstract idea of the video it captures like obstacles and road conditions, and communicate this information to other vehicles through the Internet of Vehicles to achieve self-driving. 
Moreover, emergent semantic communication is not limited to processing visual data, the semantic information can also be expressed in the form of tactile signals using vibrating bracelets. In this way, it can be utilized in real-time exoskeletons robot control or navigation for visually impaired people. 
In the future, we can even combine five-dimensional communication involving sight, smell, taste, touch, and hearing to create an immersive experience or even Metaverse. This integration of all senses can be used in healthcare, enabling physicians to command a telerobot at the patient’s location, allowing remote surgery with full MAR and haptic feedback. In the education and training area, teachers will not only have visual access to the learners but also gain a sense of their movements when engaging in tasks that require precise motor skills. This capability enables teachers to provide real-time corrections and guidance to optimize learning outcomes.
In addition to the small message size and generalization ability mentioned in the article, emergent semantic communication can also be adopted to improve communication reliability by leveraging semantic understanding to predict missing or corrupted information and fill gaps in the message context. 

Several challenges still exist in implementing emergent semantic communication. Firstly, there is a lack of systematic analysis regarding the impact of adding message error rates during the training process. While the experimental results in this paper demonstrate the enhancement of channel uncertainty immunity, the optimal error rate averages, and its potential improvement still lack mathematical analysis. Additionally, although our experimental results align with previous studies by Kucinski \emph{et al.}'s conclusion that an appropriate message error rate promotes the emergence of an ideal communication protocol \cite{Kucinski2021}, we are not yet sure about the mechanism or the mathematical proof behind it. Our initial hypothesis is that the appropriate perturbations induced by message error rate can facilitate the exploration of a wider range of possibilities.

Secondly, the existing research on emergent semantic communication primarily focuses on experimental studies, and it lacks a concise theoretical framework to guide its implementations in different scenarios. This framework should address critical aspects such as determining optimal strategies for selecting message length, vocabulary size, and neural network architectures that align with real-world requirements. Furthermore, the absence of quantitative metrics to measure performance is also a noteworthy limitation.

Finally, the scalability of emergent semantic communication in computing networks for tasks like MAR is not sufficiently explored. This paper has shown that under a specific task, two agents are able to develop an efficient and distributed communication protocol. However, as the number of agents increases, the complexity of emergent communication systems grows significantly. And coordinating and managing communication among a large number of agents might lead to inefficiencies or breakdowns in the communication process. Thus, scalability tests need to be conducted and modifications to the current training framework need to be made.

\section{Conclusion}
In this article, we proposed a basic idea and framework for emergent semantic communication for MAR tasks. Using discrete messages to communicate the abstract idea of the original video stream, this framework possesses high communication efficiency and can generalize to unseen concepts. Experimental analysis has revealed the significant advantages of our system compared to data-oriented or feature-oriented communications. To better model the real environment, we take into account different message error rates in our training process. The experiment results demonstrate that an appropriate error rate while training can improve message error immunity and even raise the accuracy of the system. At last, we discussed the opportunities and challenges in this area, providing directions for future research.

\bibliographystyle{IEEEtran} 
\bibliography{IEEEabrv,bib}

% \textbf
% {Ruxiao Chen} is currently a junior student at Shandong University. He is also an active participant in the seminar of the Shandong Key Laboratory of Wireless Communication Technologies. Despite being early in his academic journey, he has already made several contributions to the field. he has engaged in research projects and discussions with peers and professors, always seeking to expand his knowledge and hone his skills. His research interest lies in exploring the intersection of AI and edge computing, as well as how machine learning can be leveraged to solve complex problems in real-world settings.

% \textbf
% {Shuaishuai Guo}(Senior Member, IEEE) received the B.E and Ph.D. degrees in communication and information systems from the School of Information Science and Engineering, Shandong University, Jinan, China, in 2011 and 2017, respectively. He visited the University of Tennessee at Chattanooga (UTC), USA, from 2016 to 2017. He worked as a postdoctoral research fellow at King Abdullah University of Science and Technology (KAUST), Saudi Arabia from 2017 to 2019. Now, he is working as a full professor of Shandong University. His research interests include 6G communications and machine learning.

\end{document}